\documentclass[amsmath,amssymb,12pt,superscriptaddress,nofootinbib]{revtex4}

\usepackage{graphicx}
\usepackage{dcolumn}
\usepackage{bm}
\usepackage{color}

\newcommand{\dd}{\mathrm{d}}

\newcommand{\del}{\partial}
\newcommand{\ee}{{\rm e}}

\newcommand{\cm}{{\rm cm}}

\newcommand{\ds}{d_{\rm S}}

\newcommand{\dob}{d_{\rm O}}

\definecolor{DarkBlue}{rgb}{0,0,0.7} 

\definecolor{DarkRed}{rgb}{0.65,0,0}

\begin{document}
\baselineskip5.5mm
\thispagestyle{empty}

{\baselineskip0pt
\leftline{\baselineskip14pt\sl\vbox to0pt{
               \hbox{\it Yukawa Institute for Theoretical Physics} 
              \hbox{\it Kyoto University}
               \vss}}
\rightline{\baselineskip16pt\rm\vbox to20pt{
            \hbox{YITP-12-73}
            \hbox{ICRR-Report 625-2012-14}
\vss}}%
}

\author{Chul-Moon Yoo}\email{yoo@yukawa.kyoto-u.ac.jp}
\affiliation{
Yukawa Institute for Theoretical Physics, Kyoto University
Kyoto 606-8502, Japan
}

\author{Ryo Saito}\email{rsaito@yukawa.kyoto-u.ac.jp}
\affiliation{
Yukawa Institute for Theoretical Physics, Kyoto University
Kyoto 606-8502, Japan
}

\author{Yuuiti Sendouda}\email{sendouda@cc.hirosaki-u.ac.jp}
\affiliation{
Graduate School of Science and Technology, 
Hirosaki University, Hirosaki, Aomori 036-8561, Japan
}

\author{Keitaro Takahashi}\email{keitaro@sci.kumamoto-u.ac.jp}
\affiliation{
Faculty of Science, Kumamoto University, 
2-39-1, Kurokami, Kumamoto 860-8555, Japan
}

\author{Daisuke Yamauchi}\email{yamauchi@icrr.u-tokyo.ac.jp}
\affiliation{
Institute for Cosmic Ray Research, 
University of Tokyo, Kashiwa 277-8582, Japan
}

\title{Femto-lensing due to a Cosmic String}

\begin{abstract}
We consider the femto-lensing due to a cosmic string. 
If a cosmic string with the deficit angle 
$\Delta\sim 100$ [femto-arcsec] $\sim10^{-18}$ [rad] exists around 
the line of sight to a gamma-ray burst, 
we may observe characteristic interference patterns 
caused by gravitational lensing
in the energy spectrum of the gamma-ray burst. 
This ``femto-lensing'' event was first proposed 
as a tool to probe small mass primordial black holes. 
In this paper, we propose use of the femto-lensing 
to probe cosmic strings with extremely small tension. 
Observability conditions and the event rate are discussed. 
Differences between the cases of a point mass and 
a cosmic string are presented. 
\end{abstract}

\maketitle
\pagebreak

\section{Introduction}
\label{sec:intro}
Cosmic strings are line-like topological defects that are likely to emerge
through phase transitions with spontaneous symmetry breaking%
~\cite{Kibble:1976sj,Vilenkin-Shellard,Hindmarsh:1994re,Perivolaropoulos:2005wa}.
Another possibility to produce cosmic strings has been pointed
out in the context of super-string theory and their properties
are quite similar to those of field theoretic cosmic strings, 
except for the fact that inter-commuting probability between 
strings 
can be much lower than 
1~\cite{Sarangi:2002yt,Jones:2003da,Copeland:2003bj,Dvali:2003zj,Kachru:2003sx}
(see \cite{Polchinski:2004ia,Polchinski:2004hb,Davis:2005dd,Copeland:2009ga,Sakellariadou:2009ev,Ringeval:2010ca,Majumdar:2005qc,Copeland:2011dx}
for recent reviews).
Usually, cosmic strings are characterized by the string tension $\mu$. 
The geometry around an infinite straight string is locally identical
to that of flat spacetime, but globally 
it corresponds to a conical spacetime with
a deficit angle~\cite{Vilenkin:1981zs},
\begin{equation}
	\Delta =\frac{8\pi G\mu}{c^4}\,,
\end{equation}
where $G$ is Newton's gravitational constant and $c$ is the speed of light. 
One typical effect of the deficit angle is 
the formation of a 
double image of a light source 
located behind the string~\cite{Vilenkin:1981zs,Gott:1984ef}. 
We call all such effects which are caused by 
gravitational fields of a cosmic string 
gravitational lensing due to a cosmic string.

For ordinary field-theoretic strings, 
since the relevant 
energy scale is given by the symmetry breaking scale $E_{\rm SB}$\,, 
the string tension can be estimated as 
$G\mu/c^4\sim E_{\rm SB}^2/E_{\rm Pl}^2$\,, 
where $E_{\rm Pl}$ is the Planck energy. 
For the Grand Unification scale, 
we have $E_{\rm SB}\sim 10^{16}{\rm GeV}$ or 
equivalently $\Delta\sim 10^{-6}$\,. 
By contrast, the 
tension of cosmic strings from symmetry breaking 
along a supersymmetric flat direction in the potential 
may have much smaller tension. 
In this scenario, 
there can exist two typical scales which determine 
the string tension. 
One is the supersymmetry breaking scale, which can be of the order of TeV, 
and the other is the cutoff scale $\sim E_{\rm Pl}$. 
Such strings may have
the deficit angle given by $10^{-18}\lesssim\Delta\lesssim 10^{-6}$~\cite{Kawasaki:2011dp,Cui:2007js,Barreiro:1996dx,Freese:1995vp}. 
The effective four-dimensional tension of
cosmic super-strings strongly depends on the details 
of the compactification and the inflationary scenario. 
In the case of the KKLMMT scenario~\cite{Kachru:2003sx}, 
the predicted tension 
of cosmic superstrings is expected to lie in the range 
$10^{-12}\lesssim\Delta\lesssim 10^{-6}$~\cite{Sarangi:2002yt,Jones:2002cv,Jones:2003da}.

So far, many attempts have been made to search 
for the signature of cosmic strings in various 
observations including the cosmic microwave background (CMB), 
gravitational waves and gravitational lensing.
The CMB anisotropy spectrum has excluded cosmic strings with 
$ \Delta \gtrsim 10^{-6} $ from the dominant energy components of 
the universe \cite{Battye:2010xz,Yamauchi:2010ms,Urrestilla:2011gr,Dvorkin:2011aj}. 
Non-detection of gravitational waves from cosmic string loops 
also rules out the string tension 
$ \Delta \sim 10^{-5} $ \cite{Abbott:2009ws,Abbott:2009rr}.

Gravitational lensing phenomena could in principle serve as 
more direct evidence for cosmic strings~\cite{Huterer:2003ze,
Oguri:2005dt,Mack:2007ae,Kuijken:2007ma} 
although 
none have been detected yet \cite{Sazhin:2003cp,Agol:2006fb,Sazhin:2006kf}.
A recent search for lensed galaxy pairs found no 
evidence for the presence of long straight cosmic strings 
with $ \Delta > 7.5 \times 10^{-6} $ 
out to redshifts greater than $ 0.6 $ \cite{Christiansen:2010zi}.
Non-detection of characteristic variability of quasars due to 
the crossing of cosmic strings constrains lighter strings with 
$ 10^{-12} < \Delta < 10^{-8} $ down to the level of 
$ \Omega_\mathrm{cs} = 0.01 $ \cite{Tuntsov:2010fu}, where 
$ \Omega_\mathrm{cs} $ is the average density of cosmic strings 
in the units of the critical density. 
A more model-dependent but interesting bound on the 
local abundance of the strings with much lower tension 
$ 10^{-15} < \Delta < 10^{-9} $ could be obtained by 
considering the variability of Galactic stars and pulsar 
timing \cite{Pshirkov:2009vb}.

Forecasts for future constraints on the string tension include 
the B-mode polarisation of the CMB induced by straight strings with 
$ \Delta \lesssim 10^{-6} $~\cite{Urrestilla:2011gr,Dvorkin:2011aj}, weak lensing \cite{Yamauchi:2011cu,Yamauchi:2012bc},
gravitational wave bursts, the stochastic background from string loops 
with $ \Delta \lesssim 10^{-6} $~\cite{Kuroyanagi:2012wm}, 
lensing at radio frequencies by loops with 
$ \Delta \sim 10^{-8} $~\cite{Mack:2007ae} 
and the 21 cm radiation from strings with 
$ \Delta \gtrsim 10^{-9}\text{--}10^{-11} $~\cite{Khatri:2008zw}.

So far, no way to probe 
cosmic strings with $ \Delta < 10^{-16} $ 
has been proposed. 
In this paper, we propose that 
{\it femto-lensing} events of gamma-ray bursts (GRBs) 
can be used to probe the cosmological 
abundance of cosmic strings with such small tensions. 
Though the expected image separation due to a cosmic string with 
$\Delta <10^{-16}$ is too small to be angularly resolved, 
the interference between the images could induce observable characteristic 
patterns in the energy spectrum of the lensed source 
objects~\cite{1981SvAL....7..213M}. 
This effect, called femto-lensing~\cite{1992ApJ...386L...5G}, 
was first proposed as a method to probe light compact objects 
like primordial black holes (PBHs)~\cite{1992ApJ...386L...5G,1993ApJ...413L...7S,Ulmer:1994ij,Marani:1998sh} and GRBs were considered 
as the target sources. 
The interference is expected when the time delay induced by lensing 
is comparable to the inverse of the gamma-ray frequency and 
femto-lensing of GRBs 
is sensitive to compact objects with a mass range 
$10^{17}~{\rm g}\text{--}10^{20}~{\rm g}$. 
In the case of cosmic strings, as will be shown in the following sections, 
those with the deficit angle $10^{-19}\text{--}10^{-17}$ can be probed by GRBs.

Since the launch of the FERMI satellite, observational studies of 
GRBs have progressed significantly 
thanks to its unprecedented sensitivity.
Recently, a constraint on the cosmological density of compact objects 
in the mass range $5 \times 10^{17}~{\rm g}\text{--}10^{20}~{\rm g}$ was derived 
from the non-detection of femto-lensing events by the FERMI data 
in Ref.~\cite{Barnacka:2012bm}. 
Likewise, we can expect a constraint on cosmic strings with tiny 
deficit angle $10^{-19}\lesssim \Delta\lesssim10^{-17}$, 
which can hardly be probed by other observations.

The main purpose of this paper is to investigate the interference 
pattern in the energy spectrum of a GRB induced by a cosmic string 
and 
to determine how it differs 
from that for a compact object. 
The interference pattern for a compact object was discussed 
in Refs.~\cite{1993ApJ...413L...7S,Ulmer:1994ij}. 
Though typical gravitational lensing effects are often understood 
by using the geometrical optics approximation, the wave nature of 
the light is important in the case of femto-lensing. 
In the following sections, we discuss the observability of 
femto-lensing events due to cosmic strings by 
investigating wave propagation in a spacetime with 
a cosmic string~\cite{Suyama:2005ez}. 

This paper is organised as follows. 
In Sec.~\ref{wave_propagation}, 
setting a lens system composed of the observer, a source and 
a straight cosmic string, 
we present the lensed waveform. 
A derivation of the lensed waveform using the Kirchhoff integral theorem
is given in Appendix~\ref{sec:waveform}. 
The observability conditions and the event rate are 
discussed in Sec.~\ref{observability}. 
In Sec.~\ref{comparison}, we point out some differences 
between 
the case of a cosmic string and that of a point mass. 
Sec.~\ref{summary} is devoted to a summary. 

\section{Lensed Waveform}
\label{wave_propagation}

In this section, we obtain the lensed waveform of a massless field in a cosmic string spacetime and see how the oscillatory behaviour in the energy spectrum can arise due to the interference between the waves reaching the observer.

\subsection{Configuration of the Lens System}
\begin{figure}[htbp]
\begin{center}
\includegraphics[scale=0.5]{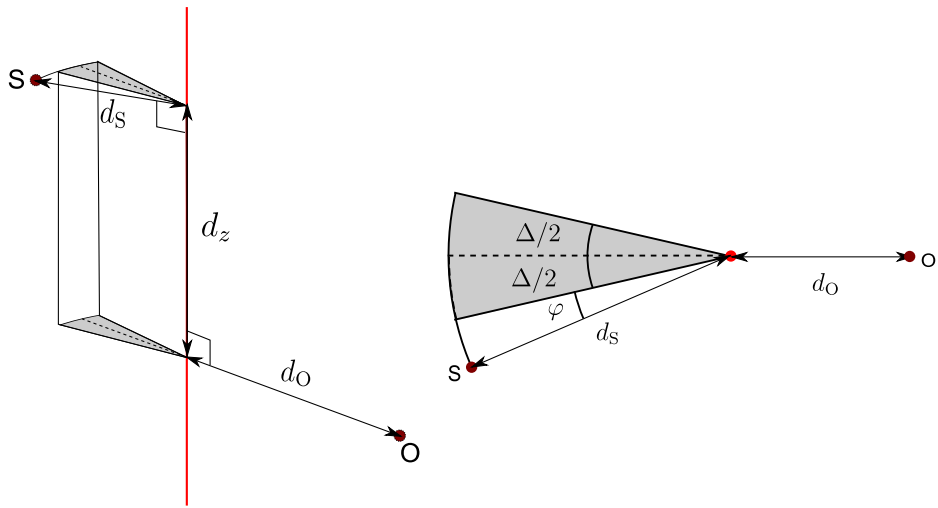}
\caption{Configuration of the lens system. The vertical straight line represents a cosmic string. The symbols O and S denote the positions of the observer and the source, respectively.}
\label{fig:config}
\end{center}
\end{figure}

We consider the spacetime with 
the deficit angle $\Delta$, 
where $\Delta$ is defined so that 
the total angle around the cosmic string 
is given by $2\pi-\Delta$. 
Let us consider the configuration of the lens system 
specified by the following quantities: 
the distances from the observer to the string $\dob$, 
from the source to the string $\ds$, 
from the observer to the source along the string direction $d_z$
and 
the angle $\varphi$ which specifies the source position as 
shown in Fig.~\ref{fig:config}. 
Hereafter, we assume $\varphi=\mathcal O(\Delta)$. 

\begin{figure}[htbp]
\begin{center}
\includegraphics[scale=0.5]{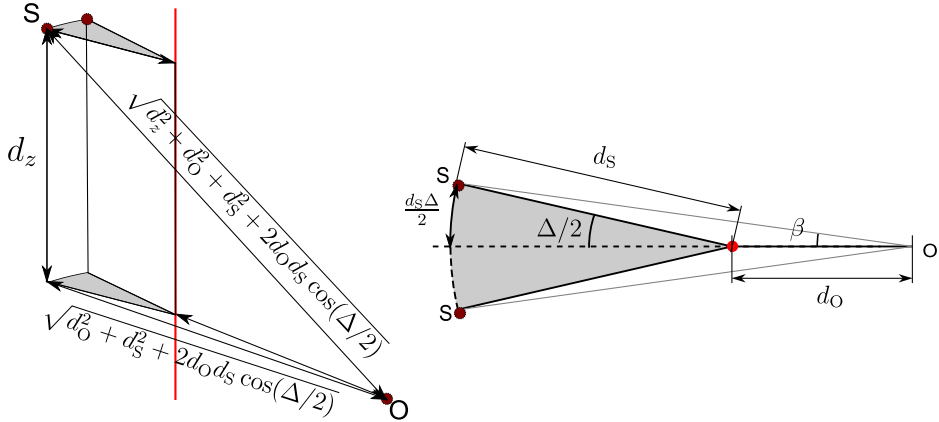}
\caption{Configuration of the lens system with $\varphi=0$.}
\label{fig:setting0}
\end{center}
\end{figure}

An important quantity characterising the wave propagation 
in the lens system is the \emph{optical path difference} $l$ defined by
\begin{equation}
l:=\frac{\dob\ds\Delta^2}{8D},
\end{equation}
where
\begin{equation}
D:=\sqrt{(\dob+\ds)^2+d_z^2}.
\end{equation}
The meaning of $ l $ is clarified by focusing on the $\varphi=0$ case, see Fig.~\ref{fig:setting0}.
Then the distance from the observer to the source
 is given by (see the left of Fig.~\ref{fig:setting0})
\begin{equation}
\sqrt{d_z^2+\dob^2+\ds^2+2\dob\ds\cos\left(\frac{\Delta}{2}\right)}
=D\left(1-
\frac{l}{D}
+\mathcal O(\Delta^4)\right).
\label{eq:distance}
\end{equation}
It is understood from Eq.~\eqref{eq:distance} that $ l $ approximately measures the shortening of the distance due to the presence of the string.

For later use we also introduce an angle
\begin{equation}
\beta:=\frac{\ds\Delta}{2\left(\dob+\ds\right)}.
\end{equation}
Again the meaning of $ \beta $ becomes clear by setting $ \varphi = 0 $:
then
 the half of the separation angle of the two images 
projected on the plane perpendicular 
to the string (see the right of Fig.~\ref{fig:setting0})
is $ \beta + \mathcal O(\Delta^2) $.

\subsection{Lensed Waveform}

Let us analyse the waveform after propagation from a stationary source to the observer.
If we neglect the tiny tilt of the polarisation vector caused by the lensing, 
the local propagation equation for  
electromagnetic waves can be 
reduced to the wave equation for the massless scalar field $\phi$: 
\begin{equation}
\left(\nabla^2 +\frac{\omega^2}{c^2}\right)\phi =
\left(\frac{\del^2}{\del r^2}+\frac{1}{r}\frac{\del}{\del r}
+\frac{1}{r^2}\frac{\del^2}{\del \theta^2}+\frac{\omega^2}{c^2}\right)\phi
=0,
\end{equation}
where $ \omega $ is the angular frequency of the mode and 
$(r,\theta,z)$ is a cylindrical coordinate system in which 
the cosmic string is on the $z$-axis. 
Since the spacetime with a deficit angle is 
locally flat, $\nabla^2$ is just 
the flat Laplacian, but the range of the azimuthal angle $\theta$ 
around the cosmic string is $[0,~2\pi-\Delta)$. 

Although we will not take cosmological expansion into account below,
since the Maxwell equations in vacuum are conformally invariant, 
we can easily apply the results to cosmological situations 
replacing Euclidean distances and the angular frequency $\omega$ 
 with 
angular diameter distances and the redshifted angular frequency $(1+z)\,\omega$, 
respectively, 
where $z$ is the redshift at the intersection of 
the string and the line of sight and $\omega$ is 
the angular frequency at the observer. 

Wave propagation in a locally flat spacetime with 
the deficit angle $\Delta$
has been fully studied in Ref.~\cite{Suyama:2005ez}.
Using appropriate approximations (see Ref.~\cite{Suyama:2005ez}\footnote{We note that notations in this paper are totally different from Ref.~\cite{Suyama:2005ez}} and Appendix~\ref{sec:waveform}),
we may solve the wave propagation to express the lensed waveform at the observer as
\begin{equation}
\phi(\dob,\ds,d_z,\varphi;\Delta,\omega)
= F(w,y)\,\phi_0(\dob,\ds,d_z,\varphi;\omega)\,,
\label{eq:lensedwave}
\end{equation}
where $ \phi_0 $ is the unlensed waveform (i.e.\ when $ \Delta = 0 $) and $ F $ is the amplification factor, respectively, given by
\begin{align}
\phi_0(\dob,\ds,d_z,\varphi;\omega)
&
:= \frac{A}{D}
   \exp\left[
    \frac{i\omega D}{c} \left(1-\frac{\dob\ds}{2D^2}\varphi^2\right)
   \right]\,, \\
F(w,y)
&
:= \exp\left[\frac{w}{2i}(1+2y)\right]
   \left\{
    1-\frac{1}{2}{\rm Erfc}\left[\sqrt{\frac{w}{2i}}(1+y)\right]
   \right\} \nonumber \\
& \qquad
   +
   \exp\left[\frac{w}{2i}(1-2y)\right]
   \left\{
    1-\frac{1}{2}{\rm Erfc}\left[\sqrt{\frac{w}{2i}}(1-y)\right]
   \right\}
\label{eq:amp}
\end{align}
with $ A $ being an arbitrary amplitude, Erfc
 the complementary error function defined by 
\begin{equation}
{\rm Erfc}(x):=\frac{2}{\sqrt{\pi}}
\int^\infty_x\ee^{-t^2}\dd t,
\end{equation}
and $ w $ and $ y $ the non-dimensional variables defined by
\begin{equation}
w
:= \frac{2\omega l}{c}
= \frac{\omega\dob\ds\Delta^2}{4cD},
\quad
y
:= \frac{\varphi\ds}{\beta(\dob+\ds)}
= \frac{2\varphi}{\Delta}. 
\label{eq:wandy}
\end{equation}
It is worth noting that 
$w$ gives roughly the ratio between the 
path difference and the wavelength,
and
$y$ gives the source position $
\varphi\,\ds$ 
on the source plane in the units of 
 $\beta\,(\dob+\ds)$.

\begin{figure}[htbp]
\includegraphics[scale=0.6]{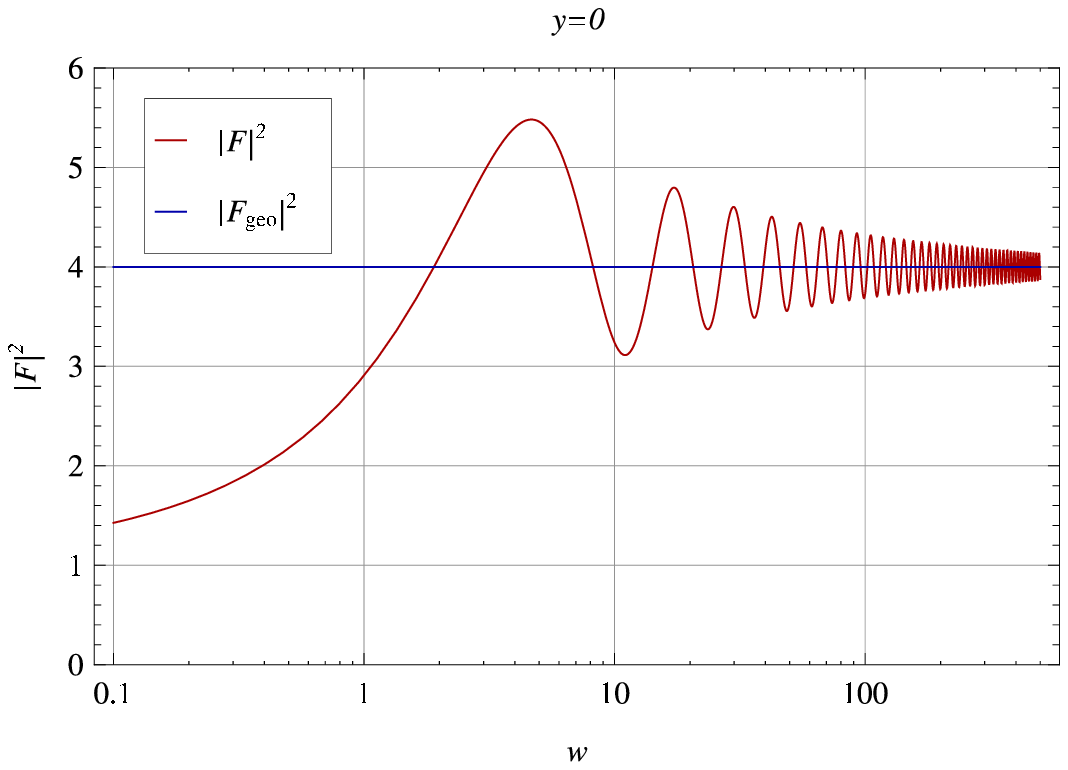}
\includegraphics[scale=0.6]{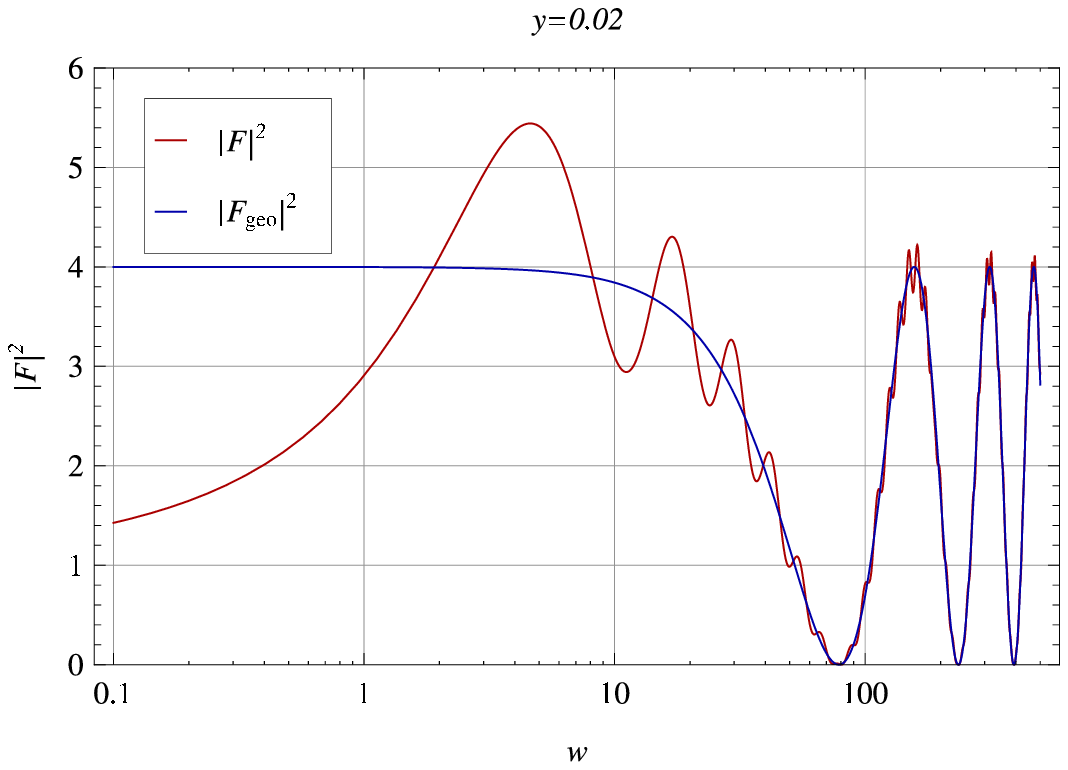}
\\
\vspace{5mm}

\includegraphics[scale=0.6]{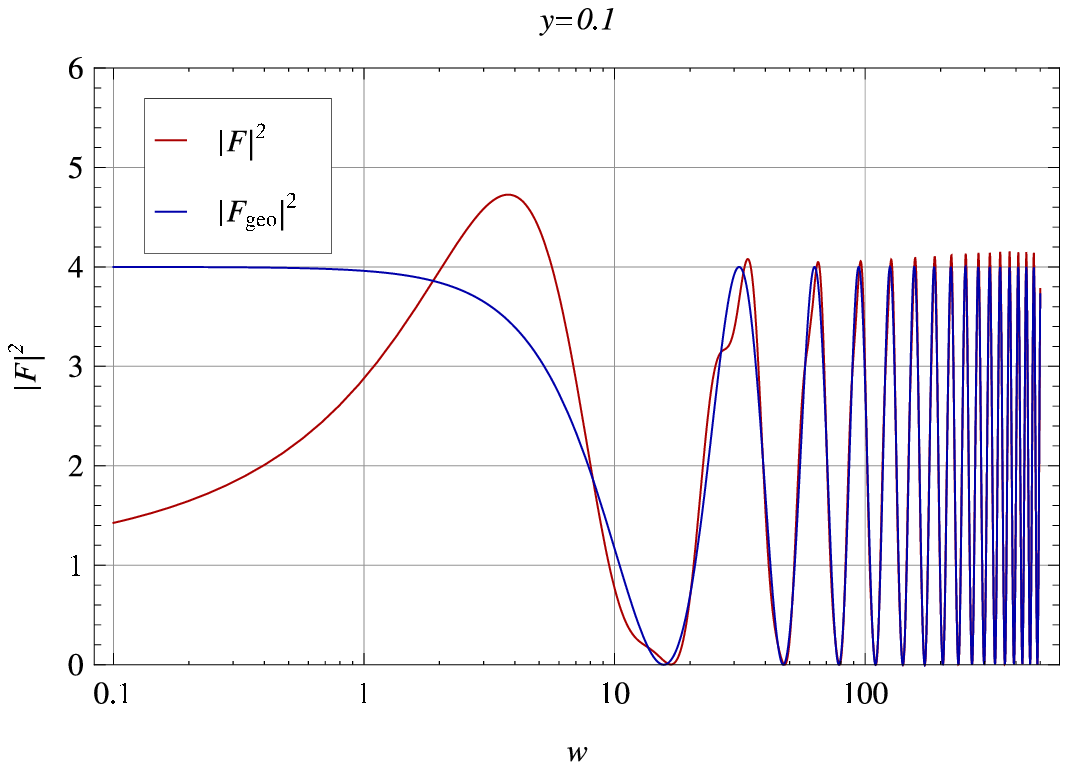}
\includegraphics[scale=0.6]{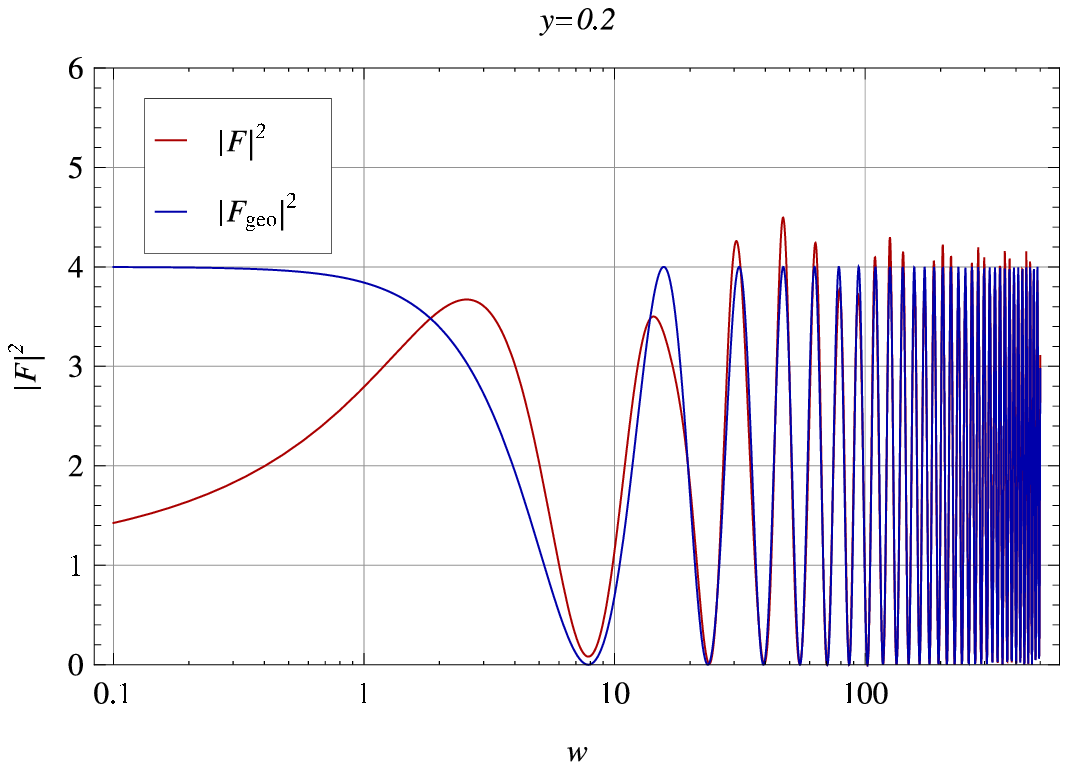}
\\
\vspace{5mm}

\includegraphics[scale=0.6]{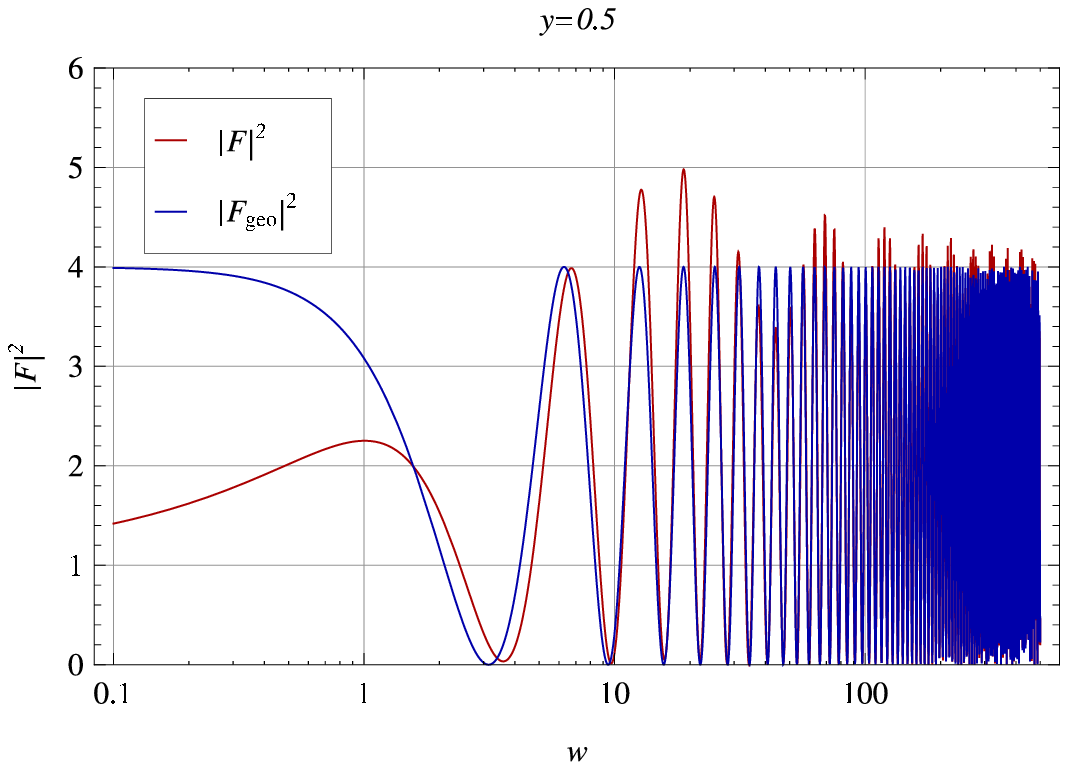}
\includegraphics[scale=0.6]{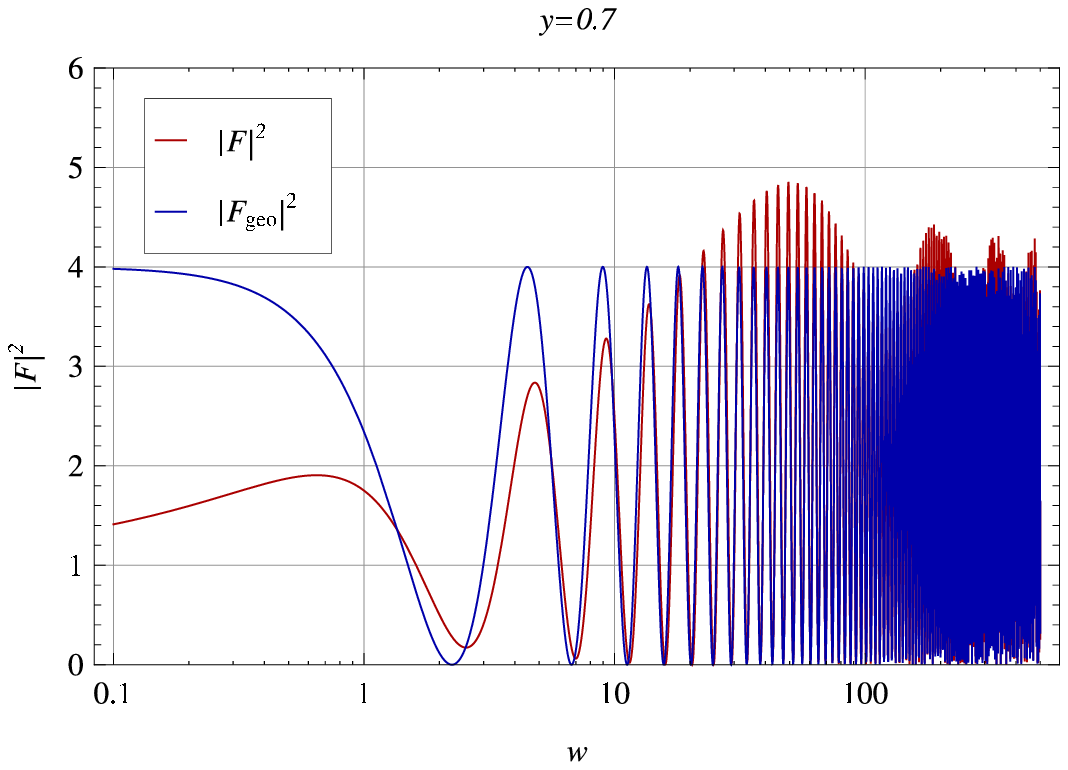}
\\
\vspace{5mm}

\includegraphics[scale=0.6]{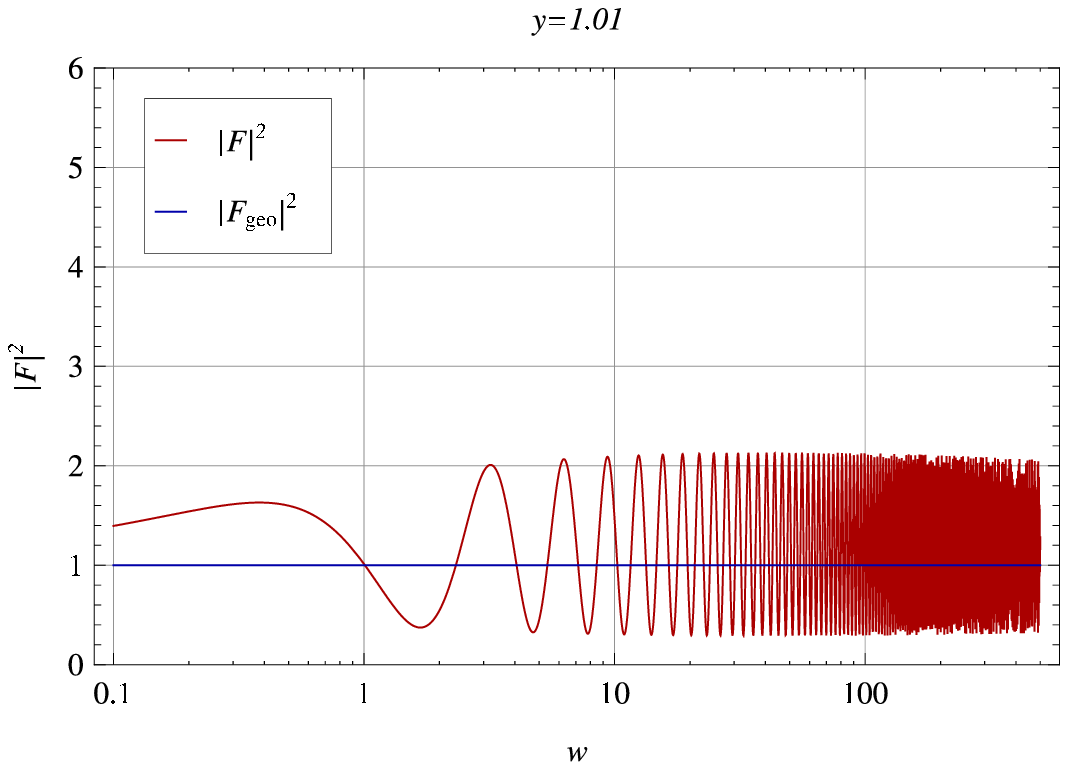}
\includegraphics[scale=0.6]{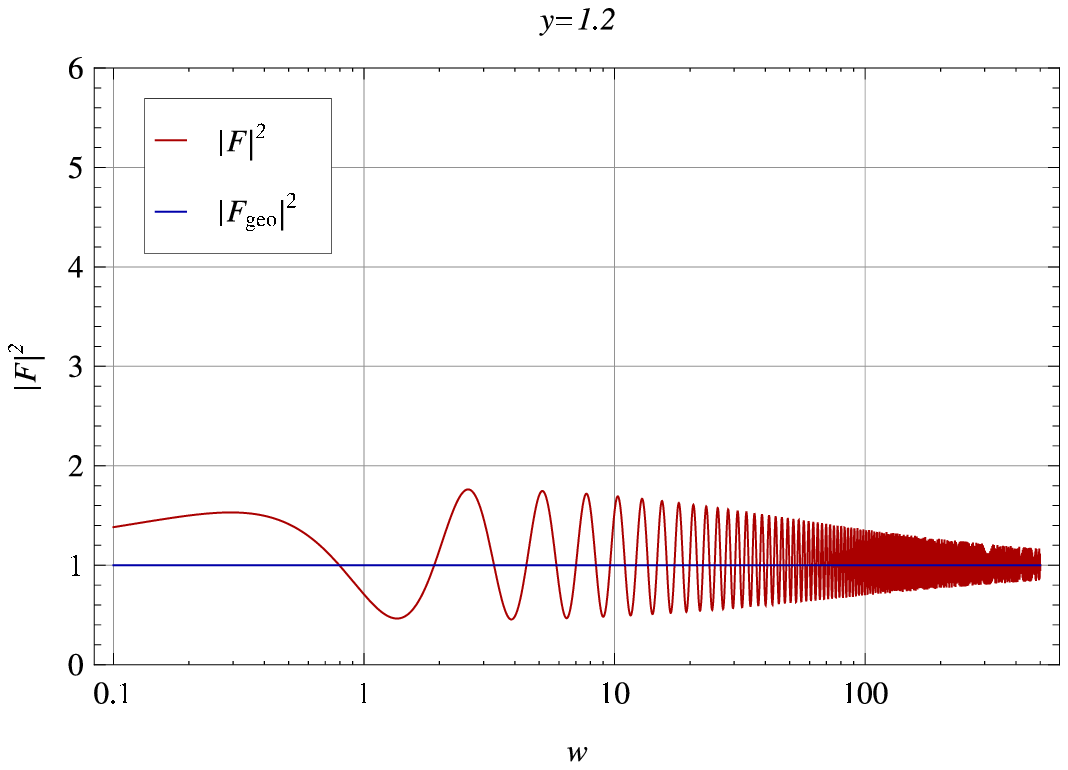}
\caption{The absolute square of the amplification factor $|F|^2$ and its geometrical optics limit $|F_{\rm geo}|^2$ as functions of $w$ for different values of $y$. 
}
\label{F2fig}
\end{figure}

We are interested in the
absolute square of $F$, 
which is observable as magnification or demagnification of 
the flux. 
It
is depicted as a function of $w$ for 
several values of $y$ in Fig.~\ref{F2fig}. 
%

The oscillatory behaviour of $ |F|^2 $ in the frequency domain is of most importance for our purpose.
In Fig.~\ref{F2fig}, the geometrical optics limit ($ w \to \infty $) of $ |F|^2 $,
\begin{equation}
\left|F_{\rm geo}\right|^2
:= \lim_{w\to\infty} \left|F\right|^2
=
\left\{
\begin{array}{lll}
1&{\rm for}& |y|>1 \\
2+2\cos(2wy)&{\rm for}& |y|<1 
\end{array}
\right.
\label{eq:stringgeo}
\end{equation}
is also depicted. 
Comparing $|F|^2$ with $|F_{\rm geo}|^2$ reveals that for most cases 
when $y>0$ the dominant mode in $|F|^2$ is the 
one that oscillates with period $ \delta w = \pi/y $. 
This is a consequence of the fact that the error function parts (those inside the curly brackets of the expression for $ F $ in Eq.~\eqref{eq:amp}) only weakly depend on $ w $.\\

To summarise this section, we have analytically obtained the lensed waveform of the massless field in the cosmic string spacetime, Eq.~\eqref{eq:lensedwave}.
The numerical plots in Fig.~\ref{F2fig} show that the absolute square of the amplification factor $ |F|^2 $ oscillates in the frequency domain with the period $ \delta\omega $ given by
\begin{equation}
\delta\omega
= \frac{\pi c}{2yl}
= \frac{4\pi D}{\dob\ds\Delta^2y}. 
\end{equation}
Using the wavelength $\lambda$, we obtain
\begin{equation}
\frac{\delta\omega}{\omega}
= \frac{\lambda}{4yl}
= \frac{2D\lambda}{\dob\ds\Delta^2y}. 
\end{equation}

\section{Observability Conditions}
\label{observability}

In this section, we apply the above analysis to gamma-ray bursts from cosmological distances to determine the conditions under which the lensing effect 
due to a string can be observable. 

\subsubsection{Detectable range of the deficit angle}

First, we evaluate the range of wavelength $\lambda$ 
within which a lensing event is detectable 
through the observation of photons with 
a fixed value of the deficit angle $\Delta$. 
At least one whole phase of 
sinusoidal oscillation should be detectable in the spectrum. 
Hence, let us focus on one period of the sinusoidal oscillation 
around the angular frequency $\omega=c\,w/2l$. 
Obviously, $w$ should satisfy
$ 2 w\,y \gtrsim  \pi $ so that at least one period can be included in the spectrum. 
Thus, a necessary condition for an observable effect is given by\footnote{
As can be seen from Fig.~\ref{F2fig}, 
the gravitational lensing effect becomes only significant when 
$w$ is larger than 1, 
so we need
$
w\gtrsim1 \Leftrightarrow \lambda \lesssim 4\pi l.
$
This inequality gives another upper bound for $\lambda$. 
}
\begin{equation}
\lambda \lesssim 8 y\,l.
\end{equation}
On the other hand, a lower bound for the wavelength is given by the 
energy resolution of the detector. 
The resolution of the detector must be 
sufficient to resolve the oscillation in the spectrum. 
We assume that the frequency resolution limit 
is given by $\alpha\omega$ with
the typical value of $ \alpha $ being $ \mathcal O(0.1) $.
Then, we obtain the following inequality as another necessary
 condition:
\begin{equation}
\alpha \lesssim \frac{\delta\omega}{\omega} = \frac{\lambda}{4yl}.
\end{equation}
Eventually, we find
\begin{equation}
\lambda \lesssim 4 y\,l \lesssim \alpha^{-1}\,\lambda
\quad
\Longleftrightarrow
\quad
4 \alpha\,y\,l \lesssim \lambda \lesssim 8 y\,l.
\label{eq:range1}
\end{equation}
This implies that only a narrow range of wavelength, with the fractional width of order $ \alpha $, is usable in probing a given configuration of the lens system.
We note that we are interested in the case $\varphi\lesssim \Delta/2$, 
or equivalently $y\lesssim1$. 

If we take $ \lambda = 10^{-9}\,\mathrm{cm} $ and $ 2 d_\mathrm O = 2 d_\mathrm S = D = 10^{28}\,\mathrm{cm} $ with the observations of GRBs in mind, we have
\begin{equation}
\alpha\,y\,\left(\frac{4 d_\mathrm O\,d_\mathrm S/D}{10^{28}\,\mathrm{cm}}\right)\,\left(\frac{\Delta}{10^{-18}}\right)^2
\lesssim \frac{\lambda}{
10^{-9}\,\mathrm{cm}}
\lesssim 
2
y\,\left(\frac{4 d_\mathrm O\,d_\mathrm S/D}{10^{28}\,\mathrm{cm}}\right)\,\left(\frac{\Delta}{10^{-18}}\right)^2. 
\label{eq:lambdarange}
\end{equation}
Thus in the case of GRBs, the typical separation angle is evaluated as $ \Delta \sim 100\,\mathrm{femto\,arcsec} $, confirming the origin of the name ``femto-lensing.''
Correspondingly, the detectable energy scale of the string tension 
 is
estimated as
\begin{eqnarray}
\left(\frac{G\mu}{c^4}\right)^{1/2}E_{\rm Pl}
&=&\left(\frac{\Delta}{8\pi}\right)^{1/2}E_{\rm Pl}
= \left(\frac{lD}{8\pi^2\dob\ds}\right)^{1/4}E_{\rm Pl}
\cr
&\sim&
5\times10^{9}{\rm GeV}\left(\frac{\lambda}{10^{-8}{\rm cm}}\right)^{1/4}
\left(\frac{4\dob\ds/D}{10^{28}{\rm cm}}\right)^{-1/4},  
\end{eqnarray}
where we have assumed $l\sim\lambda$. 

In reality, 
the wavelength of GRB photons  typically spans the finite range 
$ [\lambda_\mathrm{min},\lambda_\mathrm{max}] = [10^{-10}\,\mathrm{cm},10^{-7}\,\mathrm{cm}] $, which roughly corresponds to the energy scale from 1\,keV to 1\,MeV.
Then the detectable range of the deficit angle is found to be
\begin{multline}
2
\times 10^{-19}
y^{-1/2}
\left(\frac{\lambda_{\rm min}}{10^{-10}{\rm cm}}\right)^{1/2}
\left(\frac{4\dob\ds /D}{10^{28}{\rm cm}}\right)^{-1/2}\lesssim
\Delta 
\\
\lesssim
3\times 10^{-17}
y^{-1/2}
\left(\frac{\alpha}{0.1}\right)^{-1/2}
\left(\frac{\lambda_{\rm max}}{10^{-7}{\rm cm}}\right)^{1/2}
\left(\frac{4\dob\ds /D}{10^{28}{\rm cm}}\right)^{-1/2}. 
\end{multline}
%

\subsubsection{Source radius}
So far, it has been assumed that a 
GRB can be treated as a point source. 
However, if the source radius is 
sufficiently large, 
the interference pattern will be smeared out. 
The upper limit for the source radius can be estimated by 
considering the $y$ dependence of 
the amplification factor. 
For a fixed value of $w$, 
the square of the amplification factor 
oscillates with the period $\delta y =\pi/w$ as a function of $y$. 
The corresponding length scale on the source plane to 
this period is given by $\delta y \beta (\dob+\ds)=\pi\Delta\ds /2w$. 
If the source radius is larger than $\pi\Delta\ds /2w$, 
the interference effect is smeared out and 
the oscillation in the spectrum cannot be observed.  
Since we are mainly interested in the situation $l\sim\lambda$, we obtain $w\sim 4\pi$ and find that the source radius $r_{\rm s}$ must satisfy 
\begin{equation}
r_{\rm s}\lesssim\frac{\pi\Delta\ds}{2w}\approx
6\times 10^8\,{\rm cm}
\left(\frac{w}{4\pi}\right)^{-1}
\left(\frac{\Delta}{10^{-18}}\right)
\left(\frac{2\ds}{10^{28}{\rm cm}}\right). 
\label{eq:rs}
\end{equation}

We estimate the source radius $r_{\rm s}$ 
following Ref.~\cite{Ulmer:1994ij}. 
The appropriate linear source size is 
given by $\Gamma c \Delta t$ 
with $\Gamma$ and $\Delta t$ being the bulk Lorentz factor and the smallest variability 
time scale detected, respectively. 
Then we obtain 
\begin{equation}
r_{\rm s}\sim 3\times 10^{8}\cm 
\left(\frac{\Gamma}{100}\right)
\left(\frac{\Delta t}{ 0.1{\rm ms}} \right). 
\end{equation}
Once we fix the source radius $r_{\rm s}$, the inequality \eqref{eq:rs} 
represents the lower bound for the wave length $\lambda$ as follows:
\begin{equation}
\lambda\gtrsim 
1.5\times 10^{-10}{\rm cm}
\left(\frac{r_{\rm s}}{3\times 10^8{\rm cm}}\right)
\left(\frac{2\dob}{D}\right)
\left(\frac{\Delta}{10^{-18}}\right). 
\end{equation}
Comparing this inequality with the range of wavelength \eqref{eq:lambdarange}, 
we see that, 
although the finite source effect might be significant 
in some cases, the inequality \eqref{eq:rs} may be satisfied 
in many cases. 
We do not discuss the finite source effect in this paper.  
In a practical analysis, however, this effect may have to be taken into account
(see Ref.~\cite{1993ApJ...413L...7S} for the point mass case).

\subsubsection{
Cosmic string motion}
In the previous discussions, 
we have assumed a static configuration of the lens system. 
However, if the relative velocity between the source-observer system and 
the cosmic string is too large, 
the string would pass through the 
region in which the lensing effect is significant 
before a detector collects enough photons. 

A straight cosmic string with relativistic vertical velocity $v$ would pass through the region of interest in the time
\begin{equation}
\frac{2\dob \beta }{v}\sim
0.3\,{\rm s}\left(\frac{\Delta}{10^{-18}}\right)
\left(\frac{4\dob\ds /(\dob +\ds )}{10^{28}{\rm cm}}\right)
\left(\frac{10^{10}{\rm cm/s}}{v}\right). 
\label{timescale}
\end{equation}
A clear spectrum must be obtained within a time much shorter than this, otherwise the lens system cannot be regarded as static.
For a large value of $v$, only
a limited number of bright burst events 
would be usable to detect femto-lensing events. 
This
demand might be
compared to those in the case
of compact objects, 
which are believed to have a non-relativistic 
velocity dispersion.

\subsubsection{Cosmic string density}

Finally, we estimate how many strings in 
between the source and the observer are necessary 
in order that femto-lensing events can actually take place for a given number of GRBs. 
Here, we assume that the string network can be represented by 
a collection of straight string segments.

Suppose a spherical volume of radius $ D $, centered at the observer, contains $ N_\mathrm{cs} $ straight strings of length $ L $. 
Assuming the distances to the strings are $\sim D/2$, 
we obtain the angular scale of the length of a string as $2L/D$. 
Therefore, the solid angle 
in which the lensing effect becomes significant is given by 
$\Delta\times 2L/D$.  
If there are no overlapping regions between 
the solid angles of neighbouring strings, 
the total solid angle given by all the strings is 
$2N_{\rm cs}\Delta L/D$. 
Dividing this total solid angle by $4\pi$, 
we obtain the event
probability $P$ for a single source as 
\begin{equation}
P=\frac{2N_{\rm cs}\Delta L}{D}\times \frac{1}{4\pi}.
\end{equation}

Meanwhile, the average mass density is given in terms of $ N_\mathrm{cs} $ by
\begin{equation}
\rho_\mathrm{cs}
= \frac{N_\mathrm{cs}\,L\,\mu/c^2}{4 \pi\,D^3/3}.
\end{equation}
Introducing the density parameter of the straight strings $ \Omega_\mathrm{cs} \equiv 8 \pi\,G\,\rho_\mathrm{cs}/(3 H_0^2) $, where $ H_0 $ is the current Hubble parameter, 
we find
\begin{equation}
P
= \frac{2\Omega_{\rm cs} H_0^2 D^2}{c^2}.
\end{equation}
When we consider a source and a lens separated by cosmological distances, 
since $D\sim cH_0^{-1}$, we find $P\sim \Omega_{\rm cs}$.

In reality, a GRB consists of 
many spike emissions, and 
cosmic strings are moving with the velocity $v$. 
Even if a femto-lensing event cannot be observed 
at the first spike emission, 
it might be observed at a subsequent spike emission 
during a single GRB event.
This would in principle be possible if the GRB as a whole lasts 
longer than the crossing time scale given by \eqref{timescale} 
while there are spikes with timescales shorter than it.
Still, those spikes may not individually contain enough 
photons to provide a clear spectrum;
then we would need to collect a bunch of spikes.
Let $n_{\rm b}$ denote the mean number of such qualifying 
bunches within one GRB.
Then, the event
probability $P$ for a single 
GRB has to be multiplied by the factor $n_{\rm b}$. 
Assuming $\Omega_{\rm cs}\sim0.001$ and $n_{\rm b}\sim10$, 
we can expect roughly one femto-lensing event 
among 100 available gamma-ray burst events. 
This roughly corresponds to one detection per year with FERMI satellite 
if most of the observed gamma-ray bursts can be used for the femto-lensing search. 
Even if we do not observe such an event, 
we may obtain an observational limit on $\Omega_{\rm cs}$. 
Further detailed analysis of GRB spectra is needed 
for a more precise estimation of the event rate and 
giving observational limits on $\Omega_{\rm cs}$. 

\section{Distinction between Cosmic String and Point Mass Lens}
\label{comparison}
Cosmic strings are not the only candidates 
for a femto-lensing object, as point masses would also cause a similar effect.
It is necessary to study the differences between these two cases in order to 
determine how they can be distinguished observationally. 
Below, we shall review the femto-lensing caused by a point mass and highlight the difference between this and the case of a string.

Let us consider the point mass lens system with 
distances from the observer to a lens $D_{\rm L}$, 
from the lens to a source $D_{\rm LS}$ and 
the observer to the source $D_{\rm S}=D_{\rm L}+D_{\rm LS}$. 
Let $\eta_{\rm p}$ denote the distance between the source position 
and the intersection of the source plane and the line 
which connects the observer and the lens. 
For the point mass case, we have the following expression for 
the amplification factor~\cite{Deguchi:1986zz,1992grle.book.....S}:
\begin{equation}
\left|F\right|=\left|\ee^{\pi w_{\rm p}/4}\Gamma\left(1-\frac{iw_{\rm p}}{2}\right)
{}_1F_1\left(\frac{iw_{\rm p}}{2},1;\frac{iw_{\rm p}y_{\rm p}}{2}\right)\right|, 
\end{equation}
where $\Gamma$ and ${}_1F_1$ are the gamma function and the 
confluent hyper-geometric function, 
respectively, and 
\begin{equation}
w_{\rm p}=\frac{4\,\omega G M}{c^3},
\quad 
y_{\rm p}=\eta_{\rm p}\, \sqrt{\frac{c^2D_{\rm L}}{4GMD_{\rm LS}D_{\rm S}}}. 
\end{equation}

Considering the $\eta_{\rm p}=0$ case, 
we can find the following expressions for 
the path difference $l_{\rm p}$ and half of the 
separation angle $\beta_{\rm p}$: 
\begin{equation}
l_{\rm p}=\frac{2GM}{c^2}, 
\quad
\beta_{\rm p}=\sqrt{\frac{4GMD_{\rm LS}}{c^2D_{\rm L}D_{\rm S}}}. 
\end{equation}
These $l_{\rm p}$ and $\beta_{\rm p}$ correspond to $l$ and $\beta$ 
in the cosmic string case. 
Using these quantities, we can express 
$y_{\rm p}$ and $w_{\rm p}$ as 
\begin{equation}
w_{\rm p}
=\frac{2\omega\, l_{\rm p}}{c}, 
\quad
y_{\rm p}=\frac{\eta_{\rm p}}{\beta_{\rm p}\,(D_{\rm L}+D_{\rm LS})}.  
\end{equation}
One can see that these expressions for $y_{\rm p}$ and $w_{\rm p}$ 
are similar to those for $y$ and $w$ in the cosmic string case \eqref{eq:wandy}. 
Therefore, we identify $y_{\rm p}$ and $w_{\rm p}$ as 
$y$ and $w$, respectively, and omit 
the subscript ${\rm p}$ hereafter. 

\begin{figure}[htbp]
\begin{center}
\includegraphics[scale=1]{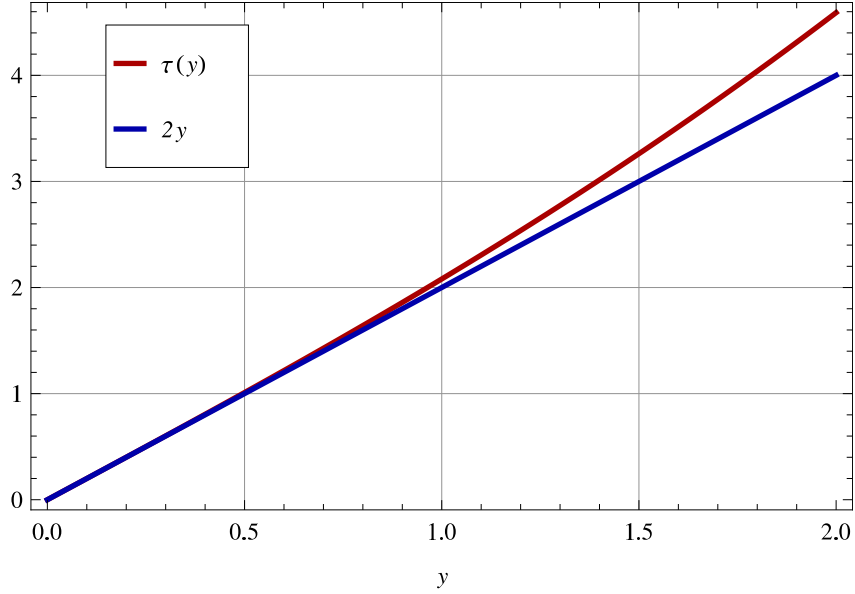}
\caption{$\tau(y)$ and $2y$. 
}
\label{fig:figtau}
\end{center}
\end{figure}

In the geometrical optics approximation ($w\rightarrow \infty$), 
we obtain 
\begin{equation}
\left|F\right|^2\rightarrow\left|F_{\rm geo}\right|^2
:=\left|\mu_+\right|+\left|\mu_-\right|
+2\sqrt{\left|\mu_+\mu_-\right|}\cos(\theta_--\theta_+),
\label{eq:pmgeo}
\end{equation}
where
\begin{eqnarray}
\mu_\pm&=&\pm \frac{1}{4}\left[\frac{y}{\sqrt{y^2+4}}
+\frac{\sqrt{y^2+4}}{y}\pm2\right], \\
\theta_+&=&w\left(\frac{1}{2x_+^2}-\ln|x_+|\right), \\
\theta_-&=&w\left(\frac{1}{2x_-^2}-\ln|x_-|\right)-\frac{\pi}{2}, 
\label{eq:caustic}
\\
x_\pm&=&\frac{1}{2}\left(y\pm\sqrt{y^2+4}\right). 
\end{eqnarray}
$\theta_--\theta_+$ can be rewritten as 
\begin{eqnarray}
\theta_--\theta_+&=&w\left[\frac{1}{2}y\sqrt{y^2+4}
+\ln\frac{\sqrt{y^2+4}+y}{\sqrt{y^2+4}-y}
\right]-\frac{\pi}{2}=:w\tau(y)-\frac{\pi}{2}. 
\end{eqnarray}
Therefore we obtain 
\begin{equation}
\cos(\theta_--\theta_+)=\sin(w\tau(y)). 
\end{equation}
It is meaningful to compare $\tau(y)$ with $2y$ because 
we have Eq.~\eqref{eq:stringgeo} for the cosmic string case in 
the geometrical optics approximation. 
As is shown in Fig.~\ref{fig:figtau}, 
$\tau(y)$ is close to $2y$ for $y\lesssim1$. 

Before comparing the cosmic string case with the point mass case, 
we compare the amplification factor with that 
of the geometrical optics approximation for each case. 
\begin{figure}[htbp]
\begin{center}
\includegraphics[scale=1]{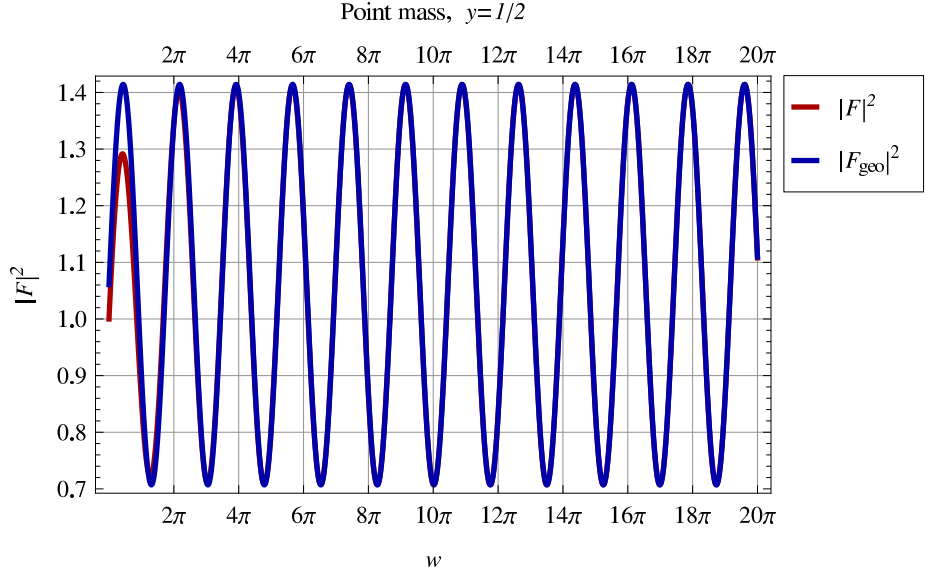}
\caption{$|F|^2$ and 
$|F_{\rm geo}|^2$ as functions of $w$ 
for the point mass case with $y=1/2$.  
}
\label{fig:Fpogeo}
\end{center}
\end{figure}
\begin{figure}[htbp]
\begin{center}
\includegraphics[scale=1]{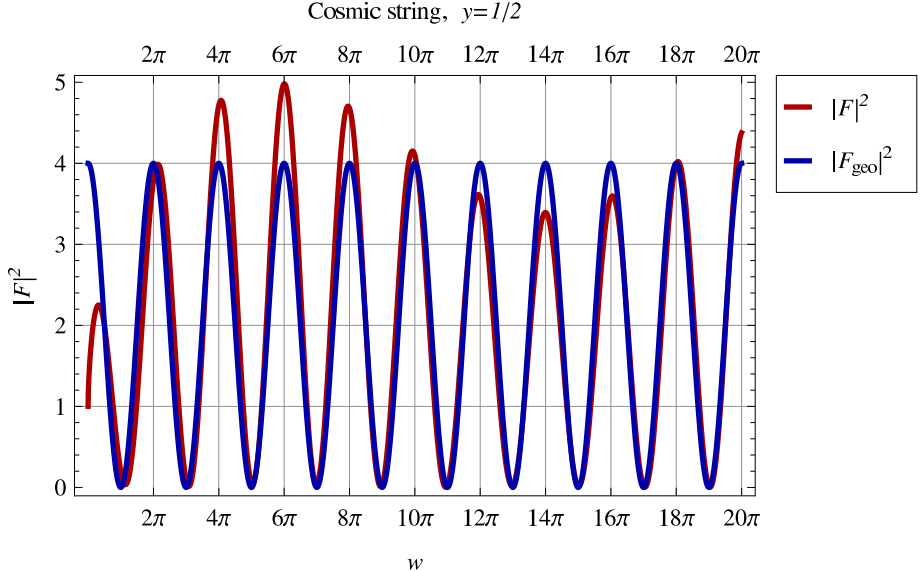}
\caption{$|F|^2$ and 
$|F_{\rm geo}|^2$ as functions of $w$ 
for the cosmic string case with $y=1/2$. 
}
\label{fig:Fstgeo}
\end{center}
\end{figure}
As is shown in Fig.~\ref{fig:Fpogeo}, 
the spectrum quickly gets closer to the geometrical optics approximation 
in the point mass case. 
Though we show only the result for $y=1/2$, 
this behaviour is a common feature for any value of $y$. 
On the other hand, this is not the case when we consider the cosmic string, 
as is shown in Fig.~\ref{fig:Fstgeo}. 
While the phase of the oscillation in the spectrum 
seems to be well approximated by 
the geometrical optics approximation, there are remarkable 
deviations in the amplitude. 
In this sense, we can conclude that the wave effect is more 
important for the cosmic string case. 

\begin{figure}[htbp]
\begin{center}
\includegraphics[scale=1]{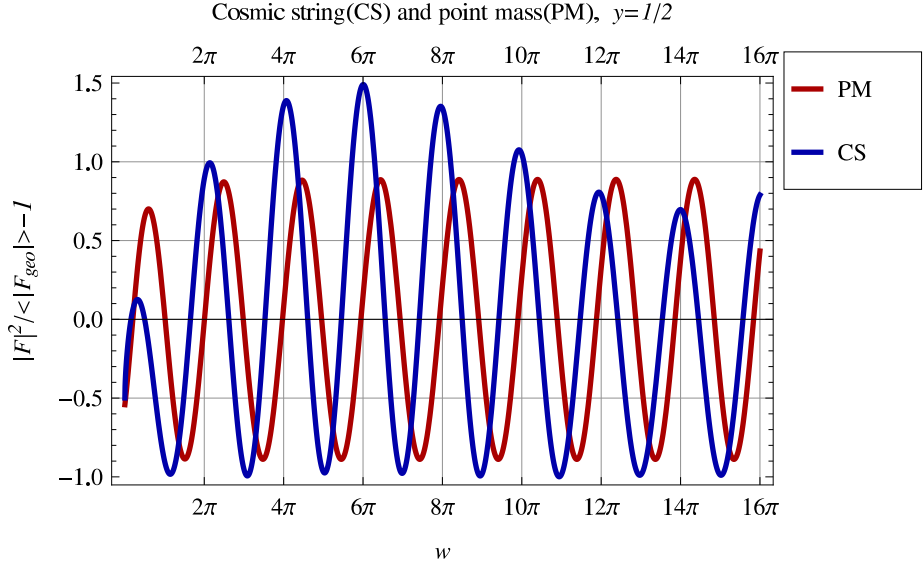}
\caption{$|F|^2/\langle |F_{\rm geo}|^2\rangle -1$ as a function of $w$ for 
the cosmic string case 
and the point mass case with $y=1/2$. 
}
\label{fig:ampdif}
\end{center}
\end{figure}

Finally, let us look at the 
differences between the cosmic string case and the point mass case. 
In order to highlight the differences, 
we focus on the following quantity:
\begin{equation}
\frac{|F|^2}{\langle |F_{\rm geo}|^2\rangle}-1, 
\end{equation}
where $\langle |F_{\rm geo}|^2\rangle$ is the 1 cycle average of the 
square of the amplification factor in the geometrical 
optics approximation. 
Using this procedure, we can extract the 
oscillating part of the spectrum. 
This quantity is plotted 
as a function of $w$ for each case in Fig.~\ref{fig:ampdif}. 

We find two remarkable differences between the two cases. 
One is the $\pi/2$ phase difference, 
the origin of which is the 
 $-\pi/2$ shift in Eq.~\eqref{eq:caustic}. 
This shift is caused by the caustic of 
one of the ray bundles~\cite{1992grle.book.....S} in the case of 
the point mass lens. 
This does not happen in the cosmic string case. 
The other difference is 
that the oscillation amplitude of $|F|^2/\langle |F_{\rm geo}|^2\rangle$ can 
exceed 1 in the cosmic string case, but 
cannot in the point mass case. 
The reason why it does not occur in the point mass case 
is obvious from Eq.~\eqref{eq:pmgeo} and the goodness of 
the geometrical optics approximation. 
These differences might be used to identify  
the lens object 
if we succeed to detect the oscillatory behaviour 
in the spectrum of a GRB. 

\section{Summary}
\label{summary}

Femto-lensing events of gamma-ray bursts (GRBs) 
due to cosmic strings have been studied. 
The detectable range of the deficit angle $\Delta$ 
was derived taking the wave nature of gamma-rays and 
the energy resolution of a detector into account. 
Assuming the range of observable wavelengths to be 
$10^{-10}{\rm cm}\text{--}10^{-7}{\rm cm}$, 
we obtain the detectable range as 
$10^{-19}\lesssim\Delta\lesssim10^{-17}$. 
Observability conditions associated with 
the source radius and the relative motion of 
the lensing system have also been discussed. 
The relative motion of a cosmic string 
gives a limitation on the duration of photon counting. 
This limitation may be tighter than that in 
the case of a compact lens object 
because of the relativistic motion
of a cosmic string. 
A limited number of bright burst events can be 
candidates for a femto-lensing event 
if the relative velocity of a cosmic string is 
comparable to the speed of light. 
The event probability $P$ for a single GRB event 
is roughly estimated as 
$P\sim n_{\rm b}\Omega_{\rm cs}$, 
where $n_{\rm b}$ and $\Omega_{\rm cs}$ are 
the mean number of available bunches of spike emissions in a GRB 
and the average density of cosmic strings 
in units of the critical density, respectively.

Two typical differences between the lensed spectrum 
in the case of a point mass lens and that of a cosmic string 
have been pointed out. 
One of these is 
a phase shift in the spectrum oscillation. 
In the point mass case, one of two ray trajectories 
experiences a caustic which causes a $-\pi/2$ phase shift, 
and this shift appears in the lensed spectrum. 
The same does not occur in the cosmic string case. 
The other difference is 
in the oscillation amplitude of the spectrum. 
In the point mass lens case, 
the oscillation amplitude does not exceed the 
mean value, while it can exceed the mean value 
in the cosmic string case because of the wave effect. 
These differences might be used to identify  
the lens object 
if we succeed to detect the oscillating behaviour 
in the spectrum of a GRB.

\section*{Acknowledgements}
We thank T.~Tanaka for helpful discussions and comments. 
We also thank J.~White for his careful reading of the manuscript and 
helpful suggestions. 
CY and RS are supported by a Grant-in-Aid through the
Japan Society for the Promotion of Science (JSPS).
This work was supported in part by MEXT Grant-in-Aid for Scientific Research on Innovative Areas No.~23740179, No.~24111710, No.~24340048 (KT)
and No.~24111701 (YS).

\appendix

\section{Derivation of the Lensed Waveform}
\label{sec:waveform}

\begin{figure}[htbp]
\begin{center}
\includegraphics[scale=0.6]{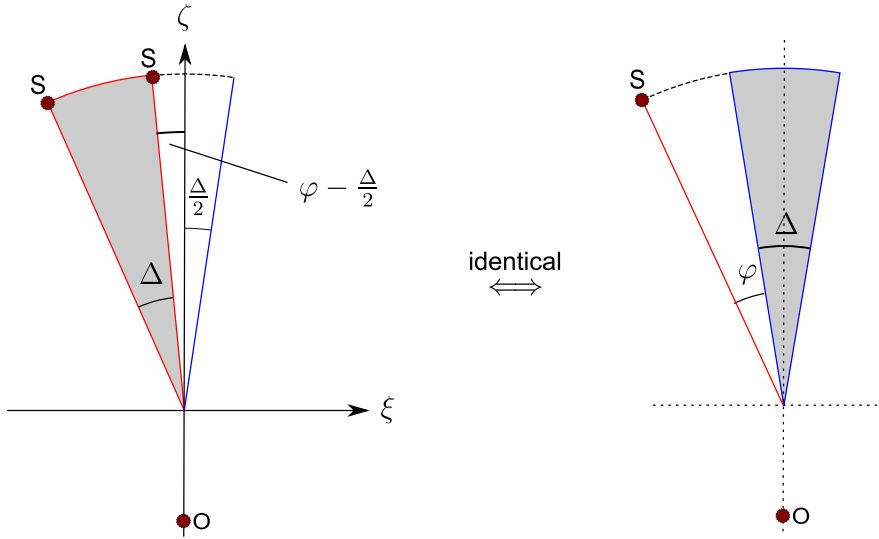}
\caption{Relation between the coordinate system $(\xi,\zeta,z)$ and 
the configuration of the lensing system is shown in the left figure. 
Another identical description which has been used in the text 
is shown in the right figure. 
}
\label{fig:app}
\end{center}
\end{figure}

We give an overview of the derivation of Eq.~\eqref{eq:lensedwave}
using the Kirchhoff integral theorem\cite{1999prop.book.....B}. 
This approach for 
the point mass lens case can be seen in Ref.~\cite{1992grle.book.....S}. 
A different approach is used in Ref.~\cite{Suyama:2005ez} 
to derive Eq.~\eqref{eq:lensedwave}. 
We use the unit $c=1$ in this
Appendix for notational simplicity. 
First, let us introduce the coordinate system $\bm \xi=(\xi,\zeta,z)$ where 
the observer and string are located at $(0,-\dob,0)$ and 
$\xi=\zeta=0$, respectively. 
For convenience, we remove the region of the deficit angle 
so that the source is just split in two 
as shown in the left of Fig.~\ref{fig:app}. 
In this coordinate system, 
the source position $\bm \xi_{\rm s}$ can be expressed in two ways. 
Let $(\xi^\pm_{\rm s}, \zeta^\pm_{\rm s}, d_z)$ denote the 
two expressions for the source position, where 
$\xi^+_{\rm s}>\xi^-_{\rm s}$. 
Then, we find 
\begin{eqnarray}
\xi^\pm_{\rm s}&=&\pm \ds\sin\left(\frac{\Delta}{2}\pm\varphi\right), \\
\zeta^\pm_{\rm s}&=&\ds\cos\left(\frac{\Delta}{2}\pm\varphi\right). 
\end{eqnarray}

Our purpose is to obtain 
an approximate solution for the wave equation given by 
\begin{equation}
(\nabla^2+\omega^2)\phi=-4\pi A \delta(\bm \xi-\bm \xi_{\rm s}).  
\end{equation}
The strategy is summarised as follows. 
First, we assume that the geometrical optics approximation is 
valid in the domain $\zeta>0$. 
Then, we calculate the waveform at the observer by applying the 
Kirchhoff integral theorem to the domain $\zeta<0$, where 
the waveform of the geometrical optics approximation on the $\zeta=0$ 
plane is used as a boundary condition. 
To evaluate the 
Kirchhoff integral, 
we use appropriate approximations associated with  
the small quantities $\varphi\sim\Delta\sim \epsilon$ and $1/(\omega D)$. 

Using the geometrical optics approximation, 
the waveform around the $\zeta=0$ plane is given by 
\begin{equation}
\left.\phi\right|_{\zeta=0}=\frac{A}{D^\pm_1}\exp(i\omega D^\pm_1) 
~~{\rm for}~~\pm \xi>0, 
\end{equation}
where 
\begin{equation}
D^\pm_1=\sqrt{(\xi-\xi^\pm_{\rm s})^2
+\left(\zeta-\zeta^\pm_{\rm s}\right)^2+
(z-d_z)^2}. 
\end{equation}
Applying the Kirchhoff integral theorem to the domain $\zeta<0$ 
and neglecting the contribution from the boundary at infinity, 
we obtain the following 
integral for the waveform at the observer:
\begin{equation}
\phi_{\rm O}:=\phi(0,-\dob,0)=-\frac{1}{4\pi}\int^\infty_{-\infty}
dz\int^\infty_{-\infty}d\xi
\left\{
\phi\frac{\del}{\del\zeta}\left(\frac{\ee^{i\omega D_2}}{D_2}\right)
-\frac{\ee^{i\omega D_2}}{D_2}
\left(\frac{\del}{\del\zeta}\phi\right)\right\}_{\zeta=0}, 
\end{equation}
where 
\begin{equation}
D_2=\sqrt{\xi^2+\left(\zeta+\dob\right)^2+z^2}. 
\end{equation}

In order to obtain an approximate form for this integral, 
we keep terms up to the order of $\epsilon^2$ in the phase 
part and up to the leading order in the amplitude. 
It is intuitively obvious that only the 
region $\xi\lesssim \epsilon D$ can give 
significant contribution to the integral. 
This fact can be justified by applying the stationary 
phase approximation to the integral over $\xi$ 
with $1/(\omega D)\ll1$. 
Taking the above discussion into account, 
the integral can be approximated as 
\begin{eqnarray}
&&\phi_{\rm O}\simeq
\frac{-i\omega A}{4\pi}
\int^\infty_{-\infty}\dd z
\left(\frac{\dob}{D_3D_5^2}+\frac{\ds}{D_3^2D_5}\right)
\exp\left[i\omega\left(D_3+D_5\right)\right]
\cr&&\times\Biggl[
\exp\left[-i\omega\left(\frac{D_5{\xi^+_{\rm s}}^2}{2D_3\left(D_3+D_5\right)}\right)
\right]
\int^\infty_0\dd\xi\exp\left[\frac{i\omega(D_3+D_5)}{2D_3D_5}\left(
\xi-\frac{D_5\xi^+_{\rm s}}{D_3+D_5}\right)^2\right]
\cr&&+
\exp\left[-i\omega\left(\frac{D_5{\xi^-_{\rm s}}^2}{2D_3\left(D_3+D_5\right)}\right)
\right]
\int^0_{-\infty}\dd\xi\exp\left[\frac{i\omega(D_3+D_5)}{2D_3D_5}\left(
\xi-\frac{D_5\xi^-_{\rm s}}{D_3+D_5}\right)^2\right]\Biggr], 
\label{eq:phio}
\end{eqnarray}
where
\begin{eqnarray}
D_3&=&\sqrt{\ds^2+(z-d_z)^2}, \\
D_5&=&\sqrt{\dob^2+z^2}. 
\end{eqnarray}
Defining $a$ and $b^\pm$ by 
\begin{eqnarray}
a&=&\frac{\omega(D_3+D_5)}{2D_3D_5}, \\
b^\pm&=&\frac{D_5\xi^\pm_{\rm s}}{D_3+D_5}, 
\end{eqnarray}
we can evaluate the integral of $\xi$ as 
\begin{equation}
\int^\infty_0\dd\xi\exp\left[ia(\pm\xi-b^\pm)^2\right]
=\sqrt{\frac{\pi i}{a}}
\left(1-\frac{1}{2} {\rm Erfc}\left(\pm\sqrt{-ia}b^\pm\right)\right). 
\end{equation}

The integral with respect to $z$ can be evaluated by using 
the stationary phase approximation.
Since the dominant contribution in the phase part comes from 
$i\omega(D_3+D_5)$, 
we write the Eq.~\eqref{eq:phio} in the following form
\begin{equation}
\phi_{\rm O}=
\int^\infty_{-\infty}\dd z
X(z)\exp[i\omega(D_3+D_5)], 
\label{eq:phio2}
\end{equation}
where 
\begin{eqnarray}
X(z)&=&
\frac{-i\omega A}{4\pi}
\left(\frac{\dob}{D_3D_5^2}+\frac{\ds}{D_3^2D_5}\right)
\cr&\times&\Biggl[
\exp\left[-i\omega\left(\frac{D_5{\xi^+_{\rm s}}^2}{2D_3\left(D_3+D_5\right)}\right)
\right]
\sqrt{\frac{\pi i}{a}}
\left(1-\frac{1}{2} {\rm Erfc}\left(\sqrt{-ia}b^+\right)\right)
\cr&+&
\exp\left[-i\omega\left(\frac{D_5{\xi^-_{\rm s}}^2}{2D_3\left(D_3+D_5\right)}\right)
\right]
\sqrt{\frac{\pi i}{a}}
\left(1-\frac{1}{2} {\rm Erfc}\left(-\sqrt{-ia}b^-\right)\right)\Biggr].  
\label{eq:x}
\end{eqnarray}
Then, we consider the stationary phase approximation 
with the phase $i\omega(D_3+D_5)$. 
Since the stationary point is 
given by 
\begin{equation}
\frac{\dd}{\dd z}(D_3+D_5)=0
\Leftrightarrow z=z':=\frac{\dob}{\dob+\ds}d_z, 
\end{equation}
Eq.~\eqref{eq:phio2} can be approximated by 
\begin{eqnarray}
\int^\infty_{-\infty}\dd z
X(z)\exp[i\omega(D_3+D_5)]
&\simeq& 
X(z')
\int^\infty_{-\infty}\dd z
\exp\left[i\omega\left\{D+\frac{(\dob+\ds)^4}{2\dob\ds D^3}
(z-z')^2\right\}\right]\cr
&=&X(z')
\sqrt{\frac{2\pi i \dob\ds D^3}{\omega(\dob+\ds)^4}}\exp[i\omega D]. 
\end{eqnarray}
Calculating $X(z')$ and simplifying the expression, 
we finally get Eq.~\eqref{eq:lensedwave}.


\begin{thebibliography}{1}

\bibitem{Kibble:1976sj}
  T.~W.~B.~Kibble,
  J.\ Phys.\ A  {\bf 9}, 1387 (1976).

\bibitem{Vilenkin-Shellard}
  A.~Vilenkin and E.~P.~S.~Shellard,
  {\it Cosmic Strings and Other Topological Defects} 
  (Cambridge University Press, Cambridge, England, 1994)

\bibitem{Hindmarsh:1994re}
  M.~B.~Hindmarsh and T.~W.~B.~Kibble,
  Rept.\ Prog.\ Phys.\  {\bf 58}, 477 (1995)
  [arXiv:hep-ph/9411342].

\bibitem{Perivolaropoulos:2005wa}
  L.~Perivolaropoulos,
  Nucl.\ Phys.\ Proc.\ Suppl.\  {\bf 148}, 128 (2005)
  [arXiv:astro-ph/0501590].

\bibitem{Sarangi:2002yt}
  S.~Sarangi and S.-H.~H.~Tye,
  Phys.\ Lett.\  B {\bf 536}, 185 (2002)
  [arXiv:hep-th/0204074].

\bibitem{Jones:2003da}
  N.~T.~Jones, H.~Stoica and S.-H.~H.~Tye,
  Phys.\ Lett.\  B {\bf 563}, 6 (2003)
  [arXiv:hep-th/0303269].

\bibitem{Copeland:2003bj}
  E.~J.~Copeland, R.~C.~Myers and J.~Polchinski,
  JHEP {\bf 0406}, 013 (2004)
  [arXiv:hep-th/0312067].

\bibitem{Dvali:2003zj}
  G.~Dvali and A.~Vilenkin,
  JCAP {\bf 0403}, 010 (2004)
  [arXiv:hep-th/0312007].

\bibitem{Kachru:2003sx}
  S.~Kachru, R.~Kallosh, A.~D.~Linde, J.~M.~Maldacena, L.~P.~McAllister and S.~P.~Trivedi,
  JCAP {\bf 0310}, 013 (2003).
  [hep-th/0308055].

\bibitem{Polchinski:2004ia}
  J.~Polchinski,
  arXiv:hep-th/0412244.

\bibitem{Polchinski:2004hb}
  J.~Polchinski,
  Int.\ J.\ Mod.\ Phys.\  A {\bf 20}, 3413 (2005)
  [AIP Conf.\ Proc.\  {\bf 743}, 331 (2005)]
  [arXiv:hep-th/0410082].

\bibitem{Davis:2005dd}
  A.-C.~Davis and T.~W.~B.~Kibble,
  Contemp.\ Phys.\  {\bf 46}, 313 (2005)
  [arXiv:hep-th/0505050].

\bibitem{Copeland:2009ga}
  E.~J.~Copeland and T.~W.~B.~Kibble,
  Proc.\ Roy.\ Soc.\ A \textbf{466}, 623--657 (2010)

\bibitem{Sakellariadou:2009ev}
  M.~Sakellariadou,
  Nucl.\ Phys.\ Proc.\ Suppl.\  {\bf 192-193}, 68 (2009)
  [arXiv:0902.0569 [hep-th]].

\bibitem{Ringeval:2010ca}
  C.~Ringeval,
  Adv.\ Astron.\  {\bf 2010}, 380507 (2010)
  [arXiv:1005.4842 [astro-ph.CO]].

\bibitem{Majumdar:2005qc}
  M.~Majumdar,
  arXiv:hep-th/0512062.

\bibitem{Copeland:2011dx}
  E.~J.~Copeland, L.~Pogosian and T.~Vachaspati,
  Class.\ Quantum Grav.\ \textbf{28}, 204009 (2011)

\bibitem{Vilenkin:1981zs}
  A.~Vilenkin,
  Phys.\ Rev.\  D {\bf 23}, 852 (1981).

\bibitem{Gott:1984ef}
  J.~R.~Gott,
  Astrophys.\ J.\  {\bf 288} (1985) 422.

\bibitem{Kawasaki:2011dp}
  M.~Kawasaki, K.~Miyamoto and K.~Nakayama,
  arXiv:1105.4383 [hep-ph].

\bibitem{Cui:2007js}
  Y.~Cui, S.~P.~Martin, D.~E.~Morrissey and J.~D.~Wells,
  Phys.\ Rev.\ D {\bf 77}, 043528 (2008)
  [arXiv:0709.0950 [hep-ph]].

\bibitem{Barreiro:1996dx}
  T.~Barreiro, E.~J.~Copeland, D.~H.~Lyth and T.~Prokopec,
  Phys.\ Rev.\ D {\bf 54}, 1379 (1996)
  [hep-ph/9602263].

\bibitem{Freese:1995vp}
  K.~Freese, T.~Gherghetta and H.~Umeda,
  Phys.\ Rev.\ D {\bf 54}, 6083 (1996)
  [hep-ph/9512211].

\bibitem{Jones:2002cv}
  N.~T.~Jones, H.~Stoica and S.~H.~H.~Tye,
  JHEP {\bf 0207}, 051 (2002)
  [hep-th/0203163].

\bibitem{Battye:2010xz}
  R.~Battye and A.~Moss,
  Phys.\ Rev.\ D {\bf 82}, 023521 (2010)
  [arXiv:1005.0479 [astro-ph.CO]].

\bibitem{Yamauchi:2010ms} 
  D.~Yamauchi, K.~Takahashi, Y.~Sendouda, C.~-M.~Yoo and M.~Sasaki,
  Phys.\ Rev.\ D {\bf 82}, 063518 (2010)
  [arXiv:1006.0687 [astro-ph.CO]].
  
  
\bibitem{Urrestilla:2011gr}
  J.~Urrestilla, N.~Bevis, M.~Hindmarsh and M.~Kunz,
  JCAP {\bf 1112}, 021 (2011)
  [arXiv:1108.2730 [astro-ph.CO]].

\bibitem{Dvorkin:2011aj}
  C.~Dvorkin, M.~Wyman and W.~Hu,
  Phys.\ Rev.\ D {\bf 84}, 123519 (2011)
  [arXiv:1109.4947 [astro-ph.CO]].

\bibitem{Abbott:2009ws}
  B.~P.~Abbott {\it et al.}  [LIGO Scientific and VIRGO Collaborations],
  Nature {\bf 460}, 990 (2009)
  [arXiv:0910.5772 [astro-ph.CO]].

\bibitem{Abbott:2009rr}
  B.~P.~Abbott {\it et al.}  [LIGO Scientific Collaboration],
  Phys.\ Rev.\ D {\bf 80}, 062002 (2009)
  [arXiv:0904.4718 [astro-ph.CO]].

\bibitem{Huterer:2003ze}
  D.~Huterer and T.~Vachaspati,
  Phys.\ Rev.\ D {\bf 68}, 041301 (2003)
  [astro-ph/0305006].

\bibitem{Oguri:2005dt}
  M.~Oguri and K.~Takahashi,
  Phys.\ Rev.\ D {\bf 72}, 085013 (2005)
  [astro-ph/0509187].

\bibitem{Mack:2007ae}
  K.~J.~Mack, D.~H.~Wesley and L.~J.~King,
  Phys.\ Rev.\ D {\bf 76}, 123515 (2007)
  [astro-ph/0702648 [astro-ph]].

\bibitem{Kuijken:2007ma}
  K.~Kuijken, X.~Siemens and T.~Vachaspati,
  Mon.\ Not.\ Roy.\ Astron.\ Soc.\ \textbf{384}, 161--164 (2008)

\bibitem{Sazhin:2003cp} 
  M.~Sazhin, G.~Longo, J.~M.~Alcal\'a, R.~Silvotti, G.~Covone, O.~Khovanskaya, M.~Pavlov, M.~Pannella, M.~Radovich and V.~Testa,
  Mon.\ Not.\ Roy.\ Astron.\ Soc.\  {\bf 343}, 353 (2003)
  [astro-ph/0302547].

\bibitem{Agol:2006fb} 
  E.~Agol, C.~J.~Hogan and R.~M.~Plotkin,
  Phys.\ Rev.\ D {\bf 73}, 087302 (2006)
  [astro-ph/0603838].

\bibitem{Sazhin:2006kf} 
  M.~V.~Sazhin, O.~S.~Khovanskaya, M.~Capaccioli, G.~Longo, M.~Paolillo, G.~Covone, N.~A.~Grogin and E.~J.~Schreier,
  Mon.\ Not.\ Roy.\ Astron.\ Soc.\  {\bf 376}, 1731 (2007)
  [astro-ph/0611744].

\bibitem{Christiansen:2010zi} 
  J.~L.~Christiansen, E.~Albin, T.~Fletcher, J.~Goldman, I.~P.~W.~Teng, M.~Foley and G.~F.~Smoot,
  Phys.\ Rev.\ D {\bf 83}, 122004 (2011)
  [arXiv:1008.0426 [astro-ph.CO]].

\bibitem{Tuntsov:2010fu} 
  A.~V.~Tuntsov and M.~S.~Pshirkov,
  Phys.\ Rev.\ D {\bf 81}, 063523 (2010)
  [arXiv:1001.4580 [astro-ph.CO]].

\bibitem{Pshirkov:2009vb} 
  M.~S.~Pshirkov and A.~V.~Tuntsov,
  Phys.\ Rev.\ D {\bf 81}, 083519 (2010)
  [arXiv:0911.4955 [astro-ph.CO]].

\bibitem{Yamauchi:2011cu} 
  D.~Yamauchi, K.~Takahashi, Y.~Sendouda and C.~-M.~Yoo,
  Phys.\ Rev.\ D {\bf 85}, 103515 (2012)
  [arXiv:1110.0556 [astro-ph.CO]].
  
\bibitem{Yamauchi:2012bc} 
  D.~Yamauchi, T.~Namikawa and A.~Taruya,
  arXiv:1205.2139 [astro-ph.CO].

\bibitem{Kuroyanagi:2012wm} 
  S.~Kuroyanagi, K.~Miyamoto, T.~Sekiguchi, K.~Takahashi and J.~Silk,
  Phys.\ Rev.\ D \textbf{86}, 023503 (2012)
  [arXiv:1202.3032 [astro-ph.CO]].

\bibitem{Khatri:2008zw} 
  R.~Khatri and B.~D.~Wandelt,
  Phys.\ Rev.\ Lett.\  {\bf 100}, 091302 (2008)
  [arXiv:0801.4406 [astro-ph]].

\bibitem{1981SvAL....7..213M}
  A.~V.~Mandzhos,
  Soviet Astronomy Letters, \textbf{7}, 213 (1981).

\bibitem{1992ApJ...386L...5G}
  A.~Gould,
  Astrophys.\ J.\ {\bf 386}, L5 (1992).

\bibitem{1993ApJ...413L...7S}
  K.~Z.~Stanek, B.~Paczynski, and J.~Goodman,
  Astrophys.\ J.\ {\bf 413}, L7 (1993).


\bibitem{Ulmer:1994ij} 
  A.~Ulmer and J.~Goodman,
  Astrophys.\ J.\  {\bf 442}, 67 (1995)
  [astro-ph/9406042].

\bibitem{Marani:1998sh}
  G.~F.~Marani, R.~J.~Nemiroff, J.~P.~Norris, K.~Hurley and J.~T.~Bonnell,
  Astrophys.\ J.\ \textbf{512}, L13 (1999)
  [astro-ph/9810391].

\bibitem{Barnacka:2012bm} 
  A.~Barnacka, J.-F.~Glicenstein and R.~Moderski,
  Phys.\ Rev.\ D \textbf{86}, 043001 (2012)

\bibitem{Suyama:2005ez}
  T.~Suyama, T.~Tanaka, and R.~Takahashi,
  Phys.\ Rev.\ {\bf D73}, 024026 (2006) [arXiv:astro-ph/0512089].

\bibitem{Deguchi:1986zz}
  S.~Deguchi and W.~D.~Watson,
  Phys.\ Rev. {\bf D34}, 1708 (1986).

\bibitem{1999prop.book.....B}
  M.~Born and E.~Wolf,
  \textit{Principles of Optics}
  (Cambridge University Press, Cambridge, England, 1999).

\bibitem{1992grle.book.....S}
  P.~Schneider, J.~Ehlers and E.~E.~Falco,
  \textit{Gravitational Lenses}
  (Springer, 1992).




\end{thebibliography}
\end{document}